\pdfoutput=1
\documentclass[12pt,a4paper]{article}
\usepackage{ifthen} % for conditional statements
\newboolean{pdflatex}
\setboolean{pdflatex}{true} % False for eps figures 

\newboolean{articletitles}
\setboolean{articletitles}{true} % False removes titles in references

\newboolean{uprightparticles}
\setboolean{uprightparticles}{false} %True for upright particle symbols

\def\paperauthors{LHCb collaboration} % Leave as is for PAPER, CONF and FIGURE
\def\paperasciititle{Precison measurement of CP violation and branching fractions in  B+/- -> KS0h+/- decays and search for the rare decay Bc+/- -> KS0 K+/-} % Set ASCII title here !! MAKE sure it's only ASCII characters !! 
\def\papertitle{Precision measurement of \CP violation and branching fractions in $B^{\pm} \to K^0_{\mathrm{S}} h^{\pm}$ $(h = \pi, K)$ decays and search for the rare decay $B_c^{\pm} \to K^0_{\mathrm{S}} K^{\pm}$} % Latex formatted title
\def\paperkeywords{{High Energy Physics}, {LHCb}} % Comma separated list
\def\papercopyright{\the\year\ CERN for the benefit of the LHCb collaboration} % new since 9/Apr/2018
\def\paperlicence{CC BY 4.0 licence}
\def\paperlicenceurl{https://creativecommons.org/licenses/by/4.0/}

\newif\ifEnableSectionTOCLinks
\EnableSectionTOCLinksfalse % deactivated

\usepackage[top=1in, bottom=1.25in, left=1in, right=1in]{geometry}

\usepackage{microtype}
\usepackage{lineno}  % for line numbering during review
\usepackage{xspace} % To avoid problems with missing or double spaces after
\usepackage{caption} %these three command get the figure and table captions automatically small

\usepackage{graphicx}  % to include figures (can also use other packages)
\usepackage{color}
\usepackage{colortbl}
\graphicspath{{./figs/}} % Make Latex search fig subdir for figures
\usepackage{amsmath} % Adds a large collection of math symbols
\usepackage{amssymb}
\usepackage{amsfonts}
\usepackage{upgreek} % Adds in support for greek letters in roman typeset
\usepackage{multirow}

\newcommand*\patchAmsMathEnvironmentForLineno[1]{%
\expandafter\let\csname old#1\expandafter\endcsname\csname #1\endcsname
\expandafter\let\csname oldend#1\expandafter\endcsname\csname
end#1\endcsname
 \renewenvironment{#1}%
   {\linenomath\csname old#1\endcsname}%
   {\csname oldend#1\endcsname\endlinenomath}%
}
\newcommand*\patchBothAmsMathEnvironmentsForLineno[1]{%
  \patchAmsMathEnvironmentForLineno{#1}%
  \patchAmsMathEnvironmentForLineno{#1*}%
}
\AtBeginDocument{%
\patchBothAmsMathEnvironmentsForLineno{equation}%
\patchBothAmsMathEnvironmentsForLineno{align}%
\patchBothAmsMathEnvironmentsForLineno{flalign}%
\patchBothAmsMathEnvironmentsForLineno{alignat}%
\patchBothAmsMathEnvironmentsForLineno{gather}%
\patchBothAmsMathEnvironmentsForLineno{multline}%
\patchBothAmsMathEnvironmentsForLineno{eqnarray}%
}

\usepackage[pdftex,
            pdfauthor={\paperauthors},
            pdftitle={\paperasciititle},
            pdfkeywords={\paperkeywords}]{hyperref}
\usepackage{hyperxmp}
\hypersetup{
    pdfcopyright={Copyright (C) \papercopyright},
    pdflicenseurl={\paperlicenceurl}
}
\usepackage[colorinlistoftodos,textsize=scriptsize]{todonotes}

\usepackage[bottom,flushmargin,hang,multiple]{footmisc}

\usepackage[all]{hypcap} % Internal hyperlinks to floats.

\usepackage{xspace} 
\usepackage{upgreek}

\def\lhcb   {\mbox{LHCb}\xspace}

\def\babar  {\mbox{BaBar}\xspace}
\def\belle  {\mbox{Belle}\xspace}
\def\belletwo {\mbox{Belle~II}\xspace}

\def\MagUp {\mbox{\em Mag\kern -0.05em Up}\xspace}

\ifthenelse{\boolean{uprightparticles}}%
{

 \def\Ppi         {\ensuremath{\uppi}\xspace}

 \def\Ppsi        {\ensuremath{\uppsi}\xspace}

 \def\PDelta      {\ensuremath{\Delta}\xspace}                 
 \def\PXi         {\ensuremath{\Xi}\xspace}                 
 \def\PLambda     {\ensuremath{\Lambda}\xspace}                 
 \def\PSigma      {\ensuremath{\Sigma}\xspace}                 
 \def\POmega      {\ensuremath{\Omega}\xspace}                 
 \def\PUpsilon    {\ensuremath{\Upsilon}\xspace}
 \let\oldPi\Pi
 \def\PPi         {\ensuremath{\oldPi}\xspace}

 \def\PB      {\ensuremath{\mathrm{B}}\xspace}                 
 \def\PD      {\ensuremath{\mathrm{D}}\xspace}                 
 \def\PJ      {\ensuremath{\mathrm{J}}\xspace}                 
 \def\PK      {\ensuremath{\mathrm{K}}\xspace}                 
 \def\Pb      {\ensuremath{\mathrm{b}}\xspace}                 
 \def\Pc      {\ensuremath{\mathrm{c}}\xspace}                 
 \def\Pd      {\ensuremath{\mathrm{d}}\xspace}                 

 \def\Pp      {\ensuremath{\mathrm{p}}\xspace}                 

 \def\Ps      {\ensuremath{\mathrm{s}}\xspace}

 \def\thebaroffset{0.0em}
}
{

 \def\Ppi         {\ensuremath{\pi}\xspace}

 \def\Ppsi        {\ensuremath{\psi}\xspace}                 
                  
 \mathchardef\PDelta="7101
 \mathchardef\PXi="7104
 \mathchardef\PLambda="7103
 \mathchardef\PSigma="7106
 \mathchardef\POmega="710A
 \mathchardef\PUpsilon="7107
 \mathchardef\PPi="7105
 \def\PB      {\ensuremath{B}\xspace}                 
 \def\PD      {\ensuremath{D}\xspace}                 
 \def\PJ      {\ensuremath{J}\xspace}                 
 \def\PK      {\ensuremath{K}\xspace}                 
 \def\Pb      {\ensuremath{b}\xspace}                 
 \def\Pc      {\ensuremath{c}\xspace}                 
 \def\Pd      {\ensuremath{d}\xspace}                 

 \def\Pp      {\ensuremath{p}\xspace}                 

 \def\Ps      {\ensuremath{s}\xspace}

 \def\thebaroffset{0.18em}
}
\newcommand{\offsetoverline}[2][\thebaroffset]{\kern #1\overline{\kern -#1 #2}}%

\makeatletter
\ifcase \@ptsize \relax% 10pt
  \newcommand{\miniscule}{\@setfontsize\miniscule{4}{5}}% \tiny: 5/6
\or% 11pt
  \newcommand{\miniscule}{\@setfontsize\miniscule{5}{6}}% \tiny: 6/7
\or% 12pt
  \newcommand{\miniscule}{\@setfontsize\miniscule{5}{6}}% \tiny: 6/7
\fi
\makeatother

\DeclareRobustCommand{\optbar}[1]{\shortstack{{\miniscule (\rule[.5ex]{1.25em}{.18mm})}
  \\ [-.7ex] $#1$}}

\def\dquark    {{\ensuremath{\Pd}}\xspace}
\def\dquarkbar {{\ensuremath{\overline \dquark}}\xspace}

\def\squark    {{\ensuremath{\Ps}}\xspace}
\def\squarkbar {{\ensuremath{\overline \squark}}\xspace}

\def\cquark    {{\ensuremath{\Pc}}\xspace}

\def\bquark    {{\ensuremath{\Pb}}\xspace}

\def\pion   {{\ensuremath{\Ppi}}\xspace}

\def\pip    {{\ensuremath{\pion^+}}\xspace}
\def\pim    {{\ensuremath{\pion^-}}\xspace}
\def\pipm   {{\ensuremath{\pion^\pm}}\xspace}

\def\kaon    {{\ensuremath{\PK}}\xspace}
\def\Kbar    {{\ensuremath{\offsetoverline{\PK}}}\xspace}

\def\KorKbar {\kern \thebaroffset\optbar{\kern -\thebaroffset \PK}{}\xspace}
\def\Kz      {{\ensuremath{\kaon^0}}\xspace}
\def\Kzb     {{\ensuremath{\Kbar{}^0}}\xspace}
\def\Kp      {{\ensuremath{\kaon^+}}\xspace}
\def\Km      {{\ensuremath{\kaon^-}}\xspace}
\def\Kpm     {{\ensuremath{\kaon^\pm}}\xspace}

\def\KS      {{\ensuremath{\kaon^0_{\mathrm{S}}}}\xspace}

\def\KL      {{\ensuremath{\kaon^0_{\mathrm{L}}}}\xspace}

\def\D       {{\ensuremath{\PD}}\xspace}

\def\DorDbar {\kern \thebaroffset\optbar{\kern -\thebaroffset \PD}\xspace}
\def\Dz      {{\ensuremath{\D^0}}\xspace}

\def\Dp      {{\ensuremath{\D^+}}\xspace}
\def\Dm      {{\ensuremath{\D^-}}\xspace}

\def\DpDm    {\ensuremath{\Dp {\kern -0.16em \Dm}}\xspace}

\def\B       {{\ensuremath{\PB}}\xspace}

\def\BorBbar {\kern \thebaroffset\optbar{\kern -\thebaroffset \PB}\xspace}
\def\Bz      {{\ensuremath{\B^0}}\xspace}

\def\Bd      {{\ensuremath{\B^0}}\xspace}

\def\BdorBdbar {\kern \thebaroffset\optbar{\kern -\thebaroffset \Bd}\xspace}
\def\Bu      {{\ensuremath{\B^+}}\xspace}
\def\Bub     {{\ensuremath{\B^-}}\xspace}
\def\Bp      {{\ensuremath{\Bu}}\xspace}
\def\Bm      {{\ensuremath{\Bub}}\xspace}
\def\Bpm     {{\ensuremath{\B^\pm}}\xspace}

\def\Bs      {{\ensuremath{\B^0_\squark}}\xspace}

\def\BsorBsbar {\kern \thebaroffset\optbar{\kern -\thebaroffset \Bs}\xspace}
\def\Bc      {{\ensuremath{\B_\cquark^+}}\xspace}
\def\Bcp     {{\ensuremath{\B_\cquark^+}}\xspace}

\def\jpsi     {{\ensuremath{{\PJ\mskip -3mu/\mskip -2mu\Ppsi}}}\xspace}

\def\Y#1S{\ensuremath{\PUpsilon{(#1S)}}\xspace}

\def\proton      {{\ensuremath{\Pp}}\xspace}

\def\Lz          {{\ensuremath{\PLambda}}\xspace}

\def\LorLbar     {\kern \thebaroffset\optbar{\kern -\thebaroffset \PLambda}\xspace}

\def\BF         {{\ensuremath{\mathcal{B}}}\xspace}
\def\BR         {\BF}

\newcommand{\decay}[2]{\ensuremath{#1\!\to #2}\xspace} 

\def\to                 {\ensuremath{\rightarrow}\xspace}

\def\CP                {{\ensuremath{C\!P}}\xspace}

\newcommand{\ACP}{{\ensuremath{{\mathcal{A}}^{\CP}}}\xspace}

\def\BcToKSK   {\decay{\Bc}{\KS\Kp}}

\def\AT#1     {\ensuremath{A_{\mathrm{T}}^{#1}}\xspace}           % 2

\def\btodds {\decay{\bquark}{\dquark\dquarkbar\squark}}
\def\btossd {\decay{\bquark}{\squark\squarkbar\dquark}}
\def\C#1      {\ensuremath{\mathcal{C}_{#1}}\xspace}                       % 9
\def\Cp#1     {\ensuremath{\mathcal{C}_{#1}^{'}}\xspace}                    % 7
\def\Ceff#1   {\ensuremath{\mathcal{C}_{#1}^{\mathrm{(eff)}}}\xspace}        % 9  
\def\Cpeff#1  {\ensuremath{\mathcal{C}_{#1}^{'\mathrm{(eff)}}}\xspace}       % 7
\def\Ope#1    {\ensuremath{\mathcal{O}_{#1}}\xspace}                       % 2
\def\Opep#1   {\ensuremath{\mathcal{O}_{#1}^{'}}\xspace}                    % 7

\newcommand{\nospaceunit}[1]{\ensuremath{\text{#1}}}       
\newcommand{\aunit}[1]{\ensuremath{\text{\,#1}}}       
\newcommand{\tev}{\aunit{Te\kern -0.1em V}\xspace}
\newcommand{\gev}{\aunit{Ge\kern -0.1em V}\xspace}
\newcommand{\mev}{\aunit{Me\kern -0.1em V}\xspace}
\newcommand{\kev}{\aunit{ke\kern -0.1em V}\xspace}
\newcommand{\ev}{\aunit{e\kern -0.1em V}\xspace}
 
\newcommand{\mevc}{\ensuremath{\aunit{Me\kern -0.1em V\!/}c}\xspace}
\newcommand{\gevc}{\ensuremath{\aunit{Ge\kern -0.1em V\!/}c}\xspace}
\newcommand{\mevcc}{\ensuremath{\aunit{Me\kern -0.1em V\!/}c^2}\xspace}
\newcommand{\gevcc}{\ensuremath{\aunit{Ge\kern -0.1em V\!/}c^2}\xspace}
\def\mm   {\aunit{mm}\xspace}

\def\mum  {\ensuremath{\,\upmu\nospaceunit{m}}\xspace}

\def\fb   {\ensuremath{\aunit{fb}}\xspace}
\def\invfb   {\ensuremath{\fb^{-1}}\xspace}

\newcommand{\chisq}{\ensuremath{\chi^2}\xspace}

\def\gsim{{~\raise.15em\hbox{$>$}\kern-.85em
          \lower.35em\hbox{$\sim$}~}\xspace}
\def\lsim{{~\raise.15em\hbox{$<$}\kern-.85em
          \lower.35em\hbox{$\sim$}~}\xspace}

\def\sPlot{\mbox{\em sPlot}\xspace}

\def\pt         {\ensuremath{p_{\mathrm{T}}}\xspace}

\def\ptot       {\ensuremath{p}\xspace}

\def\evtgen     {\mbox{\textsc{EvtGen}}\xspace}

\def\geant      {\mbox{\textsc{Geant4}}\xspace}

\def\photos     {\mbox{\textsc{Photos}}\xspace}

\def\pythia     {\mbox{\textsc{Pythia}}\xspace}

\def\tell1  {TELL1\xspace}
\def\ukl1   {UKL1\xspace}

\newcommand{\lhcborcid}[1]{\href{https://orcid.org/#1}{\hspace*{0.1em}\raisebox{-0.45ex}{\includegraphics[width=1em]{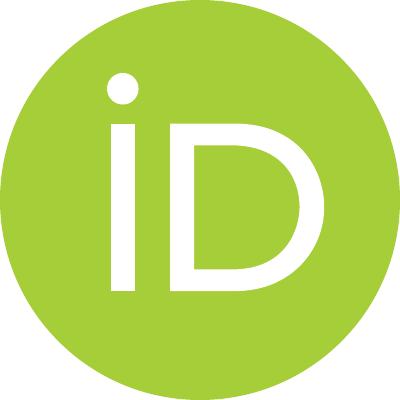}}}}

\hypersetup{
  colorlinks   = true, %Colours links instead of ugly boxes
  urlcolor     = blue, %Colour for external hyperlinks
  linkcolor    = blue, %Colour of internal links
  citecolor    = red   %Colour of citations
}

\ifEnableSectionTOCLinks
    \usepackage[explicit]{titlesec} % to change headings
    
    \let\oldcontentsline\contentsline
    \renewcommand

    \titleformat{\section}{\normalfont\Large\bf}{\hyperlink{tocsection.\thesection}{{\thesection} \parbox[t]{\dimexpr\textwidth-1pc}{#1}}}{1pc}{}

    \titleformat{\subsection}{\normalfont\bf}{\hyperlink{tocsubsection.\thesubsection}{{\thesubsection} \parbox[t]{\dimexpr\textwidth-1pc}{#1}}}{1pc}{}

    \titleformat{name=\section,numberless}[display]{}{}{0pt}{\normalfont\Huge\bfseries #1}
\fi

\usepackage{cite} % Allows for ranges in citations
\usepackage{mciteplus}
\usepackage{longtable} % only for template; not usually to be used in PAPERs

\begin{document}

%%%%%%%%%%%%%%%%%%%%%%%%%
%%%%% Title     %%%%%%%%%
%%%%%%%%%%%%%%%%%%%%%%%%%
\renewcommand{\thefootnote}{\fnsymbol{footnote}}
\setcounter{footnote}{1}

% %%%%%%% CHOOSE TITLE PAGE--------
%\onecolumn
%\input{title-LHCb-INT}
%\input{title-LHCb-ANA}
%\input{title-LHCb-CONF}
%\input{title-LHCb-FIGURE}
% ===============================================================================
% Purpose: LHCb-PAPER journal paper title page template
% Author: 
% Created on: 2010-09-25
% ===============================================================================

%%%%%%%%%%%%%%%%%%%%%%%%%
%%%%%  TITLE PAGE  %%%%%%
%%%%%%%%%%%%%%%%%%%%%%%%%
\begin{titlepage}
\pagenumbering{roman}

% Header ---------------------------------------------------
\vspace*{-1.5cm}
\centerline{\large EUROPEAN ORGANIZATION FOR NUCLEAR RESEARCH (CERN)}
\vspace*{1.5cm}
\noindent
\begin{tabular*}{\linewidth}{lc@{\extracolsep{\fill}}r@{\extracolsep{0pt}}}
\ifthenelse{\boolean{pdflatex}}% Logo format choice
{\vspace*{-1.5cm}\mbox{\!\!\!\includegraphics[width=.14\textwidth]{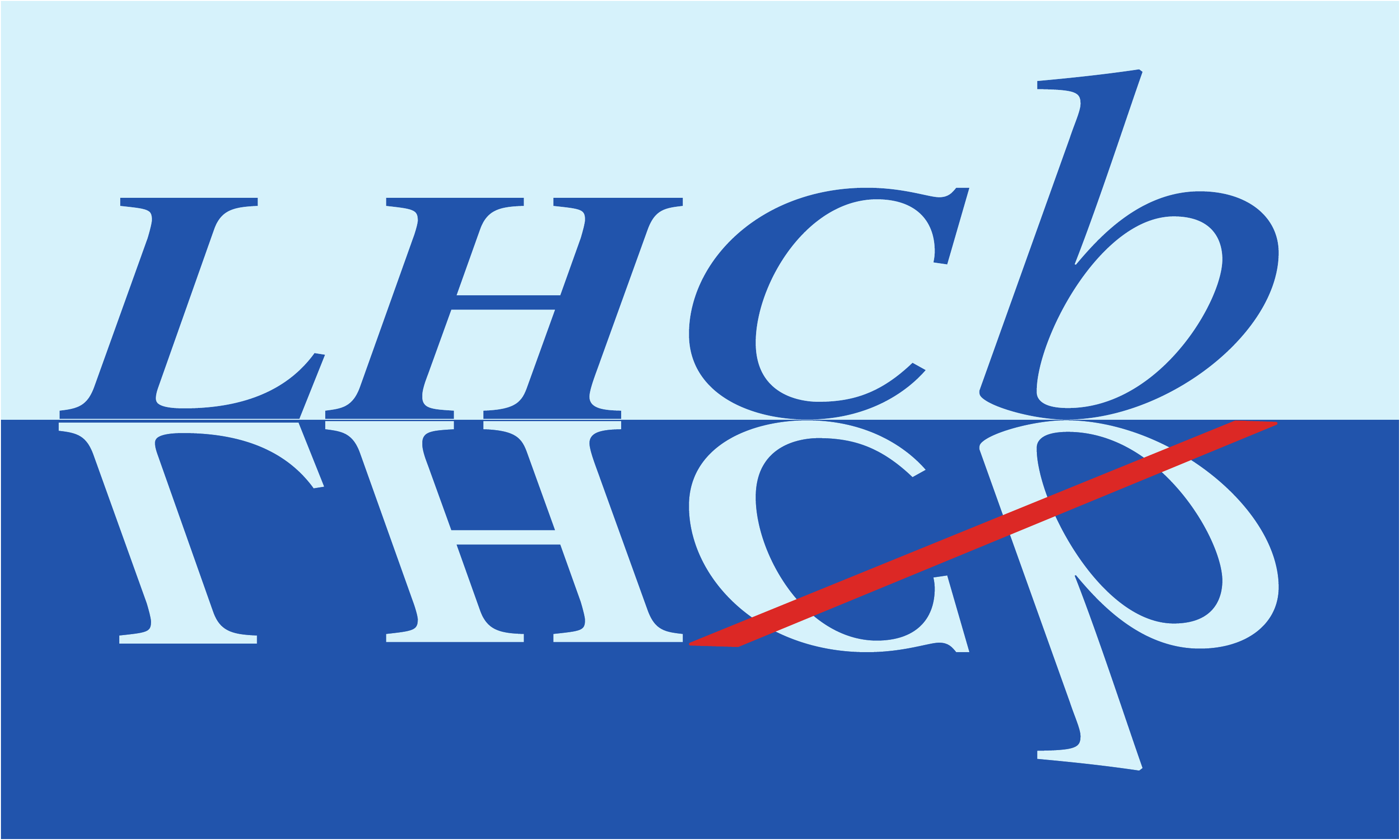}} & &}%
{\vspace*{-1.2cm}\mbox{\!\!\!\includegraphics[width=.12\textwidth]{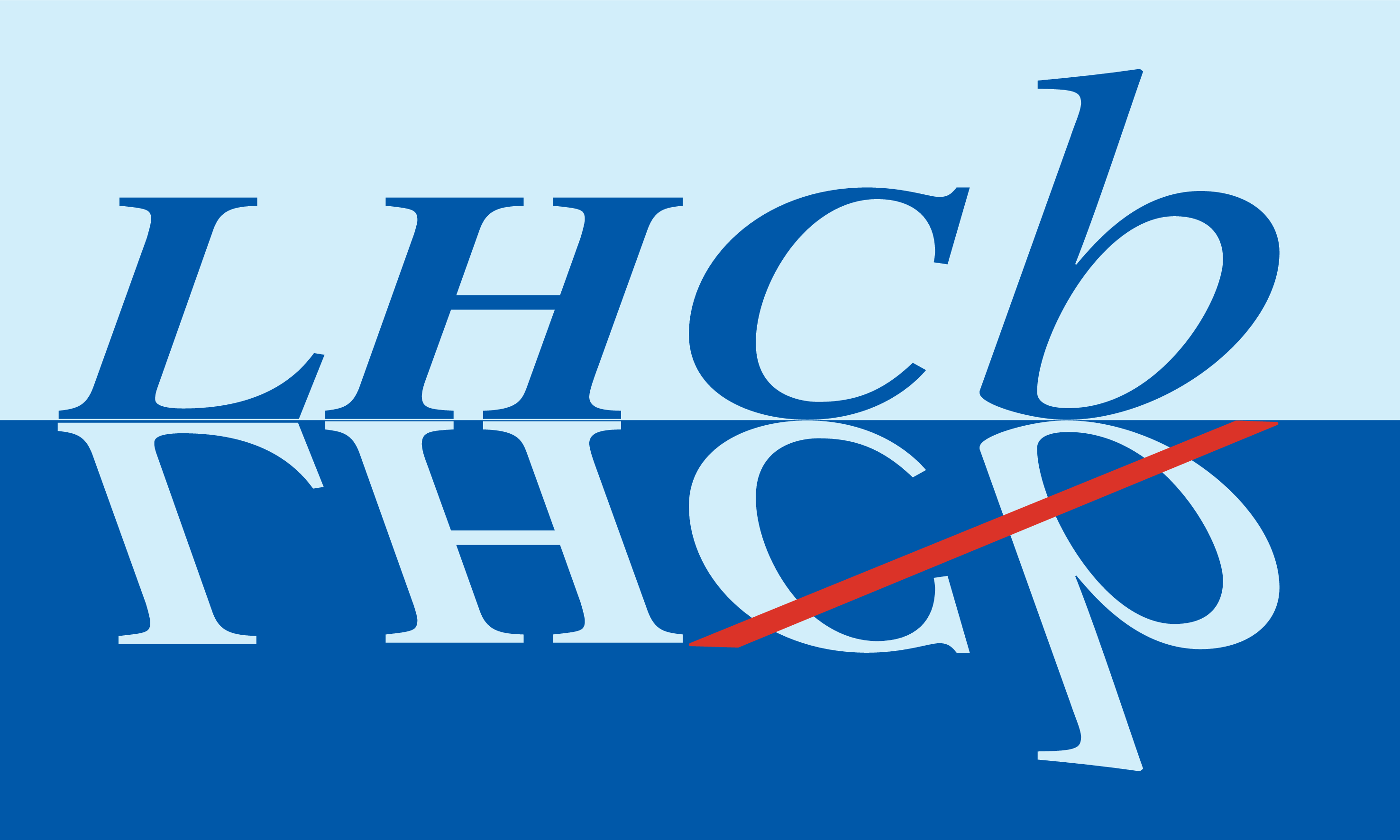}} & &}%
\\
 & & CERN-EP-2025-261 \\  % ID 
 & & LHCb-PAPER-2025-049 \\  % ID 
 & & \today \\ % Date - Can also hardwire e.g.: 23 March 2010 \today
 %& & 10 December, 2025 \\ % Date - Can also hardwire e.g.: 23 March 2010 \today
 & & \\
% not in paper \hline
\end{tabular*}

\vspace*{1.5cm}

% Title --------------------------------------------------
{\normalfont\bfseries\boldmath\huge
\begin{center}
% DO NOT EDIT HERE. Instead edit macro in main.tex to keep metadata correct
  \papertitle 
\end{center}
}

\vspace*{1.5cm}

% Authors -------------------------------------------------
\begin{center}
%In the footnote, replace 'paper' by 'Letter' in case of submission to PRL or PLB 
% Edit macro in main.tex to keep metadata correct
\paperauthors\footnote{Authors are listed at the end of this paper.}
\end{center}

\vspace{\fill}

% Abstract -----------------------------------------------
\begin{abstract}
  \noindent
% Charmless hadronic decays of charged $B$ mesons provide sensitive probes of direct $CP$ violation and loop-level flavour dynamics. 
The decay $B^{\pm} \to K^0_{\mathrm{S}} \pi^{\pm}$, with a $CP$ asymmetry expected to be close to zero in the Standard Model, is theoretically clean and sensitive to potential new physics. 
An analysis of the decays $B^{\pm} \to K^0_{\mathrm{S}} \pi^{\pm}$ and $B^{\pm} \to K^0_{\mathrm{S}} K^{\pm}$ is performed using proton-proton collision data collected by the LHCb experiment at a center-of-mass energy of~$13\tev$,  corresponding to an integrated luminosity of 5.4\,fb$^{-1}$. 
The \CP asymmetries are determined to be ${\cal A}^{CP}(B^{\pm} \to K^0_{\mathrm{S}} \pi^{\pm})=-0.028\pm 0.009\pm 0.009$ and ${\cal A}^{CP}(B^{\pm} \to K^0_{\mathrm{S}} K^{\pm})=0.118\pm 0.062 \pm 0.031$, and the branching fraction ratio is measured to be ${\cal B}(B^{\pm} \to K^0_{\mathrm{S}} K^{\pm})/{\cal B}(B^{\pm} \to K^0_{\mathrm{S}} \pi^{\pm})=0.055\pm 0.004 \pm 0.002$, where the first uncertainties are statistical and the second are systematic.
These results are the most precise measurements of these quantities to date.
A search for the rare decay $B_c^{\pm} \to K^0_{\mathrm{S}} K^{\pm}$ is also performed. No significant signal is observed, and the upper limit on the product of the branching fraction ratio ${\cal B}(B_c^{\pm} \to K^0_{\mathrm{S}} K^{\pm})/{\cal B}(B^{\pm} \to K^0_{\mathrm{S}} \pi^{\pm})$ and the fragmentation-fraction ratio $f_c/f_u$ is set to be 0.015 (0.016) at the 90\% (95\%) confidence level.
\end{abstract}

\vspace*{0.5cm}

\begin{center}
  % Submitted to
  Published in
%  JHEP /
%  Phys.~Rev.~D /
  Physical Review Letters 136, 161801 (2026). 
%  Phys.~Lett.~B /
%  Eur.~Phys.~J.~C /
  %  Nucl.~Phys.~B /
%  Chin.~Phys.~C /
%  Nature~Physics /
%  sciPost~Physics /
%  J. Instr. /
%  Instruments 
\end{center}

\vspace{\fill}

{\footnotesize 
% Edit macro in main.tex to keep metadata correct
\centerline{\copyright~\papercopyright. \href{\paperlicenceurl}{\paperlicence}.}}
\vspace*{2mm}

\end{titlepage}

%%%%%%%%%%%%%%%%%%%%%%%%%%%%%%%%
%%%%%  EOD OF TITLE PAGE  %%%%%%
%%%%%%%%%%%%%%%%%%%%%%%%%%%%%%%%

%  empty page follows the title page ----
\newpage
\setcounter{page}{2}
\mbox{~}
%\newpage
%
%% Author List ----------------------------
%%  You need to get a new author list!
%\input{LHCb_authorlist.tex}
%
%The author list for journal publications is provided by the Membership Committee shortly after 'approval to go to paper' has been given.
%%It will be made available on the page
%%\verb!http://www.physik.uzh.ch/~strauman/forMemCo/LHCb-PAPER-XXXX-XXX/! .
%It will be sent to you by email shortly after a paper number has beens assigned.
%The author list should be included already at first circulation, 
%to allow new members of the collaboration to verify whether they have been included correctly.
%Occasionally a misspelled name is corrected or associated institutions become full members.
%In that case, a new author list will be sent to you.
%In case line numbering doesn't work well after including the authorlist, try moving the \verb!\bigskip! after the last author to a separate line.
%
%
%The authorship for Conference Reports should be ``The LHCb
%  collaboration'', with a footnote giving the name(s) of the contact
%  author(s), but without the full list of collaboration names.

%\twocolumn
% %%%%%%%%%%%%% ---------

\renewcommand{\thefootnote}{\arabic{footnote}}
\setcounter{footnote}{0}

%%%%%%%%%%%%%%%%%%%%%%%%%%%%%%%%
%%%%%  Table of Content   %%%%%%
%%%%%%%%%%%%%%%%%%%%%%%%%%%%%%%%
%%%% Uncomment if desired
%\tableofcontents

\cleardoublepage

%%%%%%%%%%%%%%%%%%%%%%%%%
%%%%% Main text %%%%%%%%%
%%%%%%%%%%%%%%%%%%%%%%%%%

\pagestyle{plain} % restore page numbers for the main text
\setcounter{page}{1}
\pagenumbering{arabic}

%% Uncomment during review phase. 
%% Comment before a final submission.
%\linenumbers

%% This is the main body
%% It is useful to have a single file so comments are not missed in overleaf.
% introduction
Charmless hadronic two-body decays of \B~mesons provide powerful probes of \CP violation and rare processes, where deviations from Standard Model (SM) predictions could signal new physics in the flavour sector~\cite{Grossman:2009dw}. In particular, decays mediated by flavour-changing neutral currents 
%(FCNCs) 
offer a unique window into loop-level dynamics~\cite{Buchalla:1995vs}. The $\Bp \to \KS \pip$ decay, dominated by a single gluonic loop (penguin) amplitude with no tree-level contribution and only highly CKM-suppressed subleading terms, is among the theoretically cleanest channels in the \btodds sector~\cite{Li:2005kt,Cheng:2009cn,Cheng:2014rfa}. Within the SM, its direct \CP asymmetry is predicted to be very close to zero, making precision measurements a highly sensitive probe for potential deviations~\cite{Li:2005kt, Beneke:2001ev, Beneke:2003zv}. 
In such a theoretically clean channel, precision is the primary discriminator. Thus, advancing the experimental precision in this decay constitutes a critical step forward for testing the SM in a regime where theoretical uncertainties are minimal~\cite{Fleischer:2022rkm}.

Complementing this, the decay $\Bp \to \KS \Kp$, proceeding via the doubly CKM-suppressed \btossd transition, provides a distinct probe of penguin and annihilation amplitudes. Theoretical predictions for its \CP asymmetry exhibit strong model dependence, reflecting the intricate interplay between perturbative and nonperturbative QCD effects. Precision measurements here can discriminate between competing theoretical frameworks, such as perturbative QCD (pQCD)~\cite{Li:2005kt} and QCD factorization (QCDF)~\cite{Cheng:2009cn,Cheng:2014rfa}, and shed light on the role of poorly understood annihilation topologies. The rare decay \decay{\Bc}{\KS\Kp}, which proceeds purely via weak annihilation, offers a unique opportunity to isolate and probe this latter contribution. Experimental constraints on such processes are particularly valuable given the strong theoretical uncertainties associated with annihilation contributions in nonleptonic heavy-meson decays. 

The first studies of \decay{\Bu}{\KS\pip} decays were performed by the \babar~\cite{BaBar:2006enb} and \belle~\cite{Belle:2012dmz} collaborations. Subsequent \lhcb results, based on proton-proton ($\proton\proton$) collision data collected during 2011--2012 (Run~1), measured the ratio of branching fractions $\BR(\decay{\Bp}{\KS\Kp})/\BR(\decay{\Bp}{\KS\pip}) = 0.064\pm0.009\pm0.004$, and the \CP asymmetries $\ACP{(\decay{\Bu}{\KS\pip})} = -0.022\pm0.025\pm0.010$ and $\ACP{(\decay{\Bp}{\KS\Kp})} = -0.21\pm0.14\pm0.01$~\cite{LHCb-PAPER-2013-034}.
The \belletwo experiment reported consistent results in Ref.~\cite{Belle-II:2023ksq}.
This Letter presents updated measurements of \CP asymmetries for both the \decay{\Bu}{\KS\pip} and \decay{\Bp}{\KS\Kp} decays
with significantly improved precision.
The analysis uses a subset of Run~2 data, recorded by the \lhcb experiment between 2016 and 2018 at a center-of-mass energy of 13\tev, corresponding to an integrated luminosity of 5.4\invfb. 
Taking advantage of the same data sample and analysis strategy, a search for the pure annihilation decay \decay{\Bc}{\KS\Kp} is also performed, providing a complementary probe of annihilation amplitudes. In order to avoid experimenter’s bias, the results of the analyses were not examined until the full procedure had been finalized.

%% detector
The \lhcb detector~\cite{LHCb-DP-2008-001,LHCb-DP-2014-002} is a single-arm forward spectrometer covering the \mbox{pseudorapidity} range $2<\eta <5$,
designed for the study of particles containing \bquark or \cquark
quarks. 
The detector used to collect the data for this analysis includes a high-precision tracking system
consisting of a silicon-strip vertex detector surrounding the $pp$
interaction region~\cite{LHCb-DP-2014-001}, a large-area silicon-strip detector located upstream of a dipole magnet with a bending power of about $4\,\aunit{T\kern 0.1em m}$, and three stations of silicon-strip detectors and straw drift tubes~\cite{LHCb-DP-2017-001} placed downstream of the magnet.
The tracking system provides a measurement of the momentum, \ptot, of charged particles with a relative uncertainty that varies from 0.5\% at low momentum to 1.0\% at 200\gevc.
The minimum distance of a track to a primary $pp$ collision vertex (PV), the impact parameter, is measured with a resolution of $(15+29/\pt)\mum$, where \pt is the component of the momentum transverse to the beam, in\,\gevc.
Different types of charged hadrons are distinguished using information from two ring-imaging Cherenkov detectors~\cite{LHCb-DP-2012-003}. 
Photons, electrons and hadrons are identified by a calorimeter system consisting of scintillating-pad and preshower detectors, an electromagnetic and a hadronic calorimeter. 
The online event selection is performed by a trigger~\cite{LHCb-DP-2012-004}, which consists of a hardware stage, based on information from the calorimeter and muon systems, followed by a software stage, which applies a full event reconstruction. At the hardware stage, events are required to include a hadron with high transverse energy in the calorimeters. The software trigger requires a two- or three-track secondary vertex with a significant displacement from any PV.

Simulation is required to model the effects of the detector acceptance and the imposed selection requirements. In the simulation, $pp$ collisions are generated using \pythia~\cite{Sjostrand:2007gs,*Sjostrand:2006za} with a specific \lhcb configuration~\cite{LHCb-PROC-2010-056}. 
Decays of unstable particles are described by \evtgen~\cite{Lange:2001uf}, in which final-state radiation is generated using \photos~\cite{davidson2015photos}.
The interaction of the generated particles with the detector, and its response, are implemented using the \geant toolkit~\cite{Allison:2006ve, *Agostinelli:2002hh} as described in Ref.~\cite{LHCb-PROC-2011-006}. 

%% event selection
The \decay{\Bu}{\KS\pip} and \decay{\Bp}{\KS\Kp} candidates are formed by combining a \mbox{\decay{\KS}{\pip\pim}} candidate decay with a charged track that is identified as a pion or kaon. 
Pions from \KS decays must satisfy $\ptot>2\gevc$ and be incompatible with production at any PV. The \KS candidates are required to have $\ptot>8\gevc$, $\pt>1\gevc$, a small vertex-fit \chisq, a mass within $\pm15\mevcc$ of the known \KS mass~\cite{PDG2024}, and a significant separation of the decay vertex from all PVs, corresponding to a flight-distance significance larger than $2\sigma$. Final-state pions and kaons must satisfy $\pt>1\gevc$ and be inconsistent with PV origin. The \Bu candidates are selected with $\ptot >25\gevc$, and must have a high-quality vertex displaced from all PVs by at least 1\mm. 
Further background suppression is achieved by requiring significant displacement between the \KS decay vertex and the reconstructed \Bu vertex along the beam direction. Fiducial volume requirements are imposed to exclude \Bu candidates near detector acceptance boundaries where detection asymmetries are enhanced.

Verification of consistency between data and simulation and further background suppression is achieved, respectively, using a boosted decision tree (BDT) classifier implemented via the AdaBoost algorithm~\cite{AdaBoost} and a gradient boosted decision tree (BDTG) classifier~\cite{Breiman}, implemented via the TMVA toolkit~\cite{Hocker:2007ht,*TMVA4}. 
Both algorithms use four classes of discriminating variables: kinematic properties including the \Bu candidate \pt; vertex-quality information including the \KS and \Bu vertex-fit \chisq and the \chisq of a kinematic refit assuming the \Bu originates from a PV~\cite{Hulsbergen:2005pu}; lifetime-related information including the impact parameters of charged hadrons, \KS, and \Bu candidates as well as the \Bu flight distance from its PV; and isolation variables characterizing the \Bu candidate.
The classifiers are trained using background from the upper \Bp mass sideband in data (\mbox{$5600<m_{\Bu}<6000\mevcc$}), and signal from simulated \decay{\Bu}{\KS\pip} and \decay{\Bp}{\KS\Kp} decays. The simulation is corrected for differences with respect to data by using the \decay{\Bu}{\KS\pip} channel, which has the largest signal yield. In this channel, a loose BDT selection is applied, and the background is statistically subtracted using the \sPlot technique~\cite{Pivk:2004ty}. A gradient-boosting weighter is then used to adjust the simulated distributions to match the background-subtracted data. The following variables are considered for the matching: the \Bu candidate's \pt, $\eta$, cosine of the angle between momentum and flight direction, and decay time; the \chisq of a kinematic refit of the \Bu decay; and the track multiplicity. Owing to the limited size of the \decay{\Bu}{\KS\Kp} sample, the correction derived from the \decay{\Bu}{\KS\pip} sample is applied to both modes. 

For the final selection stage, separate BDTG classifiers are trained and applied to both the \decay{\Bu}{\KS\pip} and \decay{\Bp}{\KS\Kp} channels. 
The optimal BDTG output requirement is determined separately for \KS\pipm and \KS\Kpm samples by maximizing the signal significance $N_{\rm S}/\sqrt{N_{\rm S}+N_{\rm B}}$. The expected background yield, $N_{\rm B}$, is determined from the sideband data and extrapolated into the signal region (\mbox{$5219.34<m_{\Bu}<5339.34\mevcc$}), while the expected signal yield, $N_{\rm S}$, is obtained from simulation. To further suppress specific peaking backgrounds, a set of dedicated mass vetoes is implemented. The \decay{\Lz}{\proton \pim} veto, designed to suppress misidentified \KS candidates, is applied to both the \decay{\Bu}{\KS\pip} and \decay{\Bp}{\KS\Kp} decay modes. Additionally, exclusively for the \decay{\Bp}{\KS\Kp} channel, two further mass vetoes are introduced: one to suppress cross-feed from \decay{\Bu}{\KS\pip} decays where the pion is misidentified as a kaon, and another to reject contributions from \decay{\Bu}{\Dz \pip} decays with \decay{\Dz}{\Km \pip}. These vetoes are optimized using both data and simulation.

%% signal extraction
The total signal yields and raw charge asymmetries for \decay{\Bp}{\KS\Kp} and \decay{\Bu}{\KS\pip} decays are determined through a simultaneous unbinned extended maximum-likelihood fit to the \Bpm mass distributions in the range \mbox{$5000<m_{\Bu}<5700\mevcc$}. The fit incorporates four samples: \decay{\Bu}{\KS\pip}, \decay{\Bm}{\KS\pim}, \decay{\Bp}{\KS\Kp} and \decay{\Bm}{\KS\Km}. Each mass distribution contains four distinct components that are modeled separately: signal, cross-feed, partially reconstructed background and combinatorial background.
The signal component is parameterized using the sum of a modified Gaussian function with power-law tails on both sides (double-sided Crystal Ball, DSCB)~\cite{Skwarnicki:1986xj} and a Gaussian function, with a common peak position. The ratio of widths of the DSCB and Gaussian components, and the tail parameters of the DSCB function, are fixed to values obtained from simulation.

Due to the residual misidentification between pions and kaons after imposing particle identification (PID) requirements, a small fraction of \decay{\Bu}{\KS\pip} decays may be mistakenly reconstructed as \KS\Kp candidates, and vice versa. The cross-feed components from \decay{\Bp}{\KS\Kp} and \decay{\Bu}{\KS\pip} decays are both modeled with a DSCB function. Misidentification rates are then measured with a data-driven approach and applied as weights to the simulation sample. The cross-feed yields are constrained within statistical uncertainties to the values expected from the corresponding signal yields. Additional background contributions arise from partially reconstructed decays, predominantly from \Bz, \Bu, and \Bs mesons decaying to open-charm modes, with smaller contributions from three-body charmless decays. These backgrounds are modeled using an ARGUS function~\cite{ALBRECHT1990278} convolved with a Gaussian resolution function. The remaining combinatorial background is parameterized with a first-order polynomial function.

In the fit to data, the signal peak positions and widths are treated as free parameters, shared between \Bu and \Bm. 
The width of the ${\Bu\to\KS\Kp}$ signal is scaled relative to that of ${\Bu\to\KS\pip}$ using a simulation-derived factor. The peak shifts between signal and cross-feed components are also determined from simulation.

\begin{figure}[!tbp]
    \centering
    \includegraphics[width=0.49\linewidth]{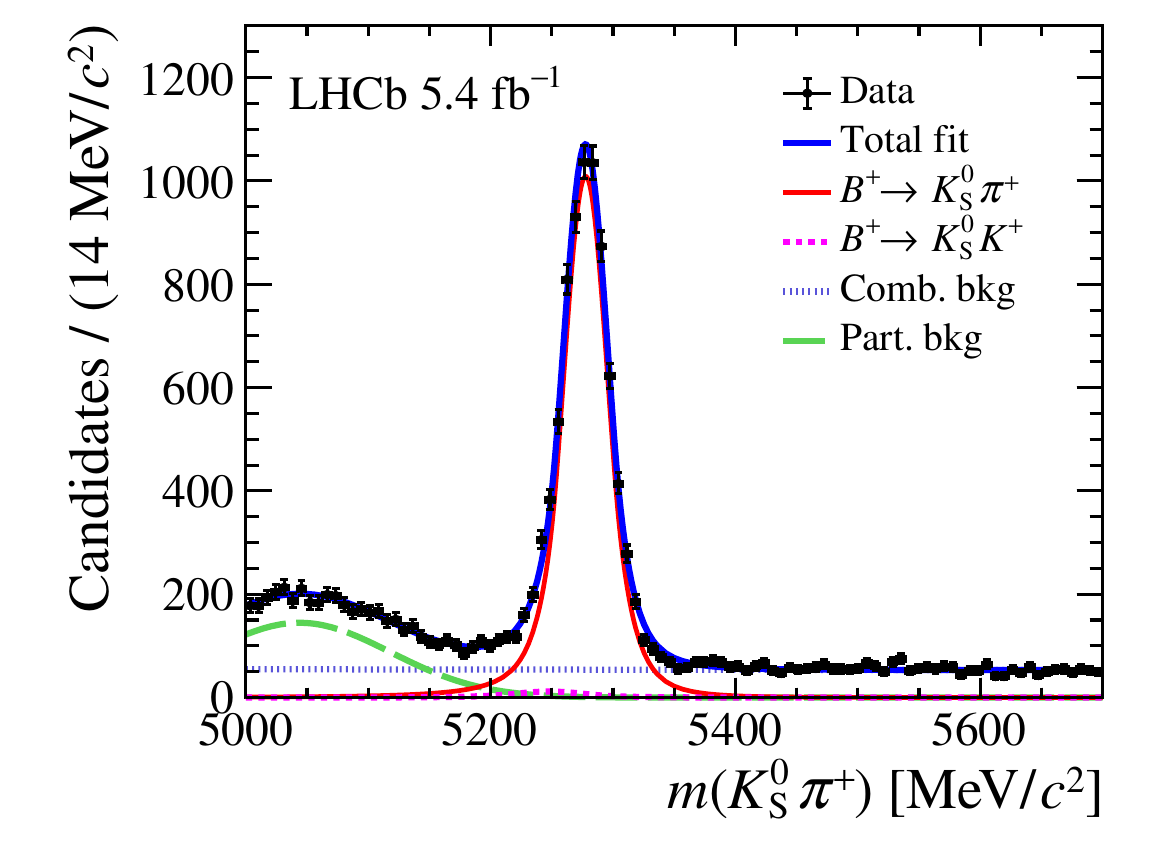}
    \includegraphics[width=0.49\linewidth]{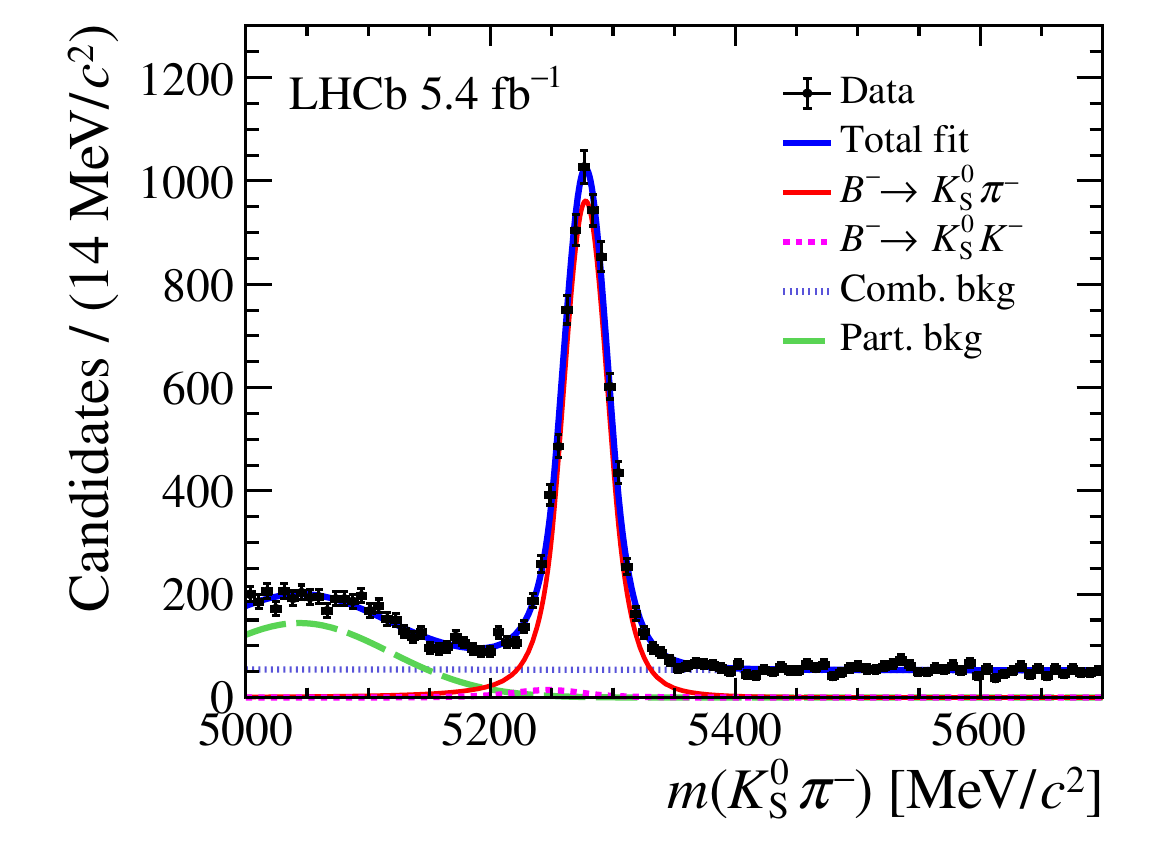}
    \includegraphics[width=0.49\linewidth]{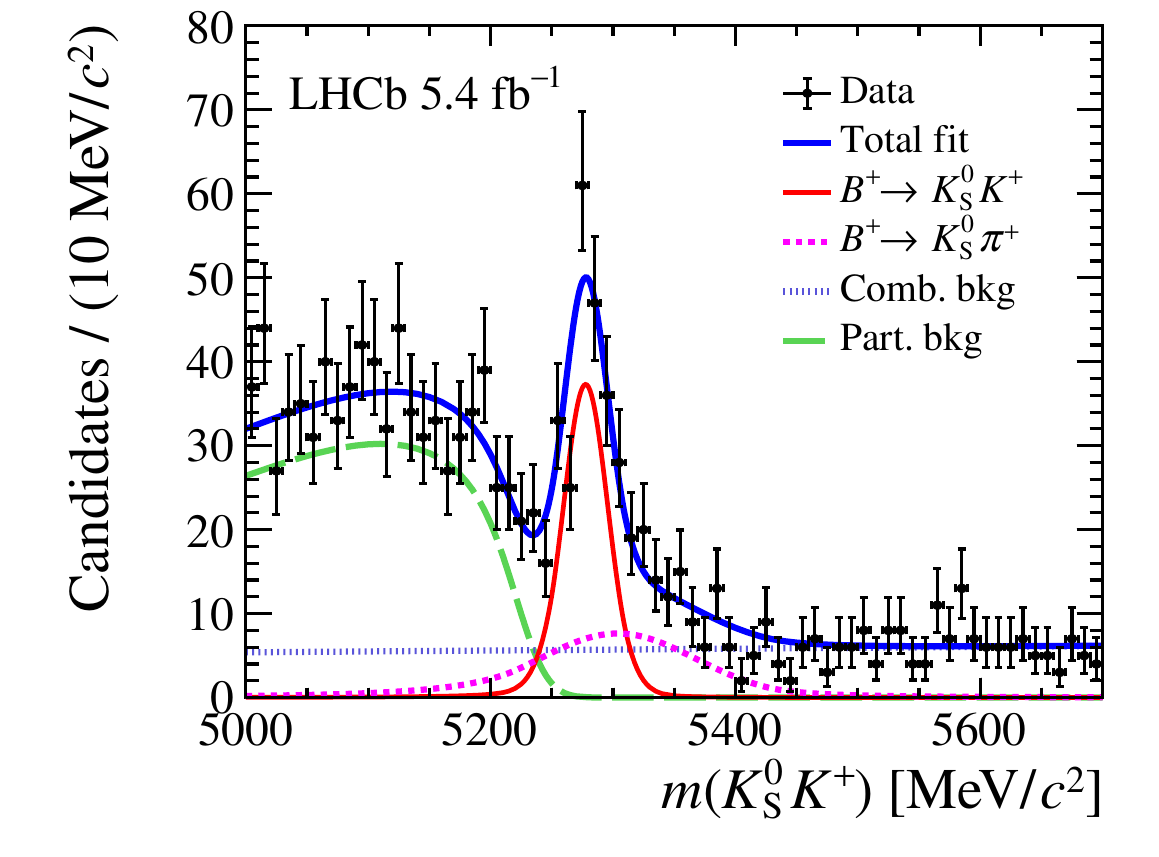}
    \includegraphics[width=0.49\linewidth]{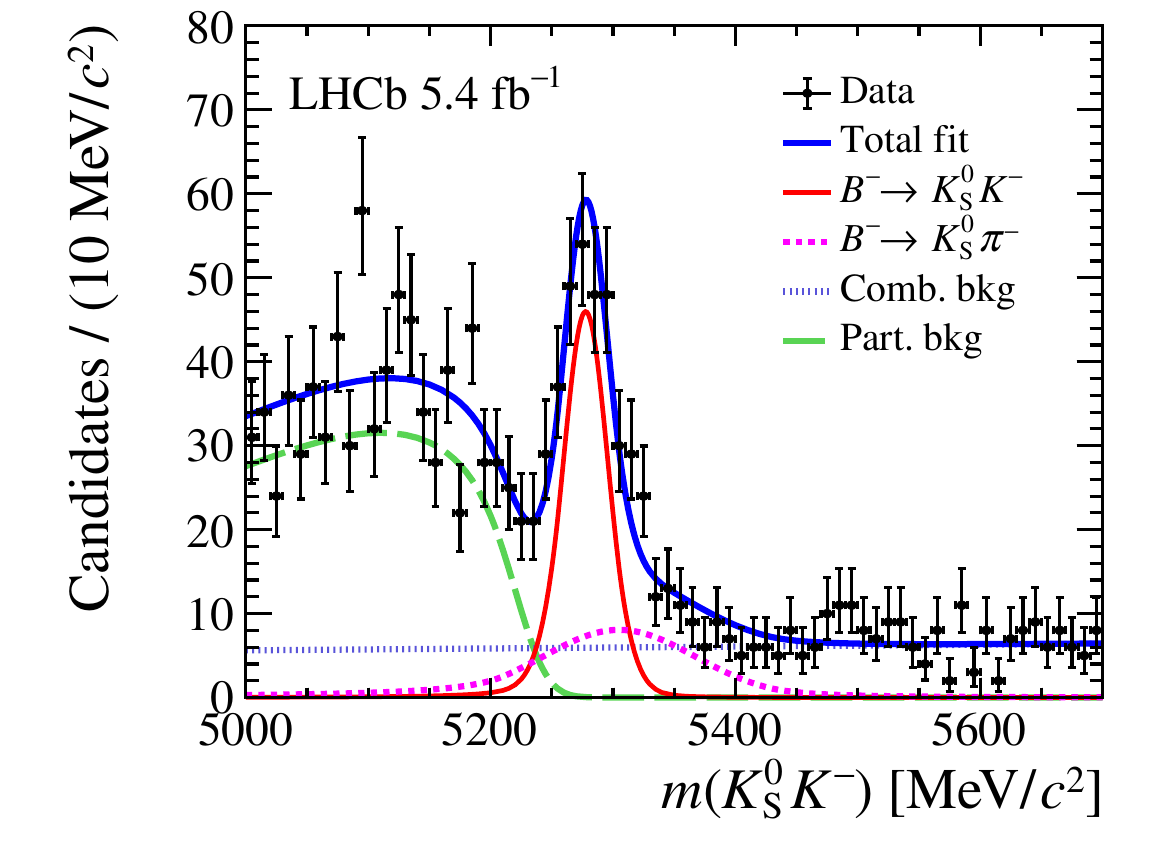}
    \caption{Mass distributions of (top left) $K^0_{\mathrm{S}} \pi^{+}$, (top right) $K^0_{\mathrm{S}} \pi^{-}$, (bottom left) $K^0_{\mathrm{S}} K^{+}$, and (bottom right) $K^0_{\mathrm{S}} K^{-}$ pairs. For visualisation, the data are presented with binned distributions using 100 bins for $K^0_{\mathrm{S}} \pi^{\pm}$ and 70 bins for $K^0_{\mathrm{S}} K^{\pm}$. The results of the simultaneous fit are also shown. }
    \label{fits}
\end{figure}

The raw charge asymmetry is defined as
   \begin{align}
    \mathcal{A}^{\textrm{raw}}_{X} & \equiv \frac{N(\Bm \to \offsetoverline{X}) - N(\Bp\to X)}{N(\Bm\to \offsetoverline{X}) + N(\Bp\to X)},
   \end{align}
where $N(\Bpm\to {\kern \thebaroffset\optbar{\kern -\thebaroffset X}\xspace})$ are the signal yields obtained from the fit. The mass distributions of the selected \decay{\Bpm}{\KS\pipm} and \decay{\Bpm}{\KS\Kpm} candidates, along with the fit projections, are shown in Fig.~\ref{fits}. The total signal yields are $N(\decay{\Bpm}{\KS\pipm})=14\,482\pm156$ and $N(\decay{\Bpm}{\KS\Kpm})=393\pm28$, with corresponding raw charge asymmetries of \mbox{$\mathcal{A}^{\textrm{raw}}_{\KS\pip} = -0.030 \pm 0.009$} and \mbox{$\mathcal{A}^{\textrm{raw}}_{\KS\Kp} = 0.109 \pm 0.062$}. 

The raw charge asymmetries for the \decay{\Bu}{\KS\pip} and \decay{\Bp}{\KS\Kp} decays are corrected for detection and production asymmetries, $\mathcal{A}^{\textrm{det+prod}}$, as well as for a small contribution from \CP violation in the neutral kaon system, $\mathcal{A}^{\textrm{\Kz}}$. The latter is assumed to be identical for both decay modes. At first order, the \CP asymmetry for the \decay{\Bu}{\KS\Kp} decay is
\begin{eqnarray}
\ACP{(\decay{\Bu}{\KS\Kp})} & \approx & \mathcal{A}^{\textrm{raw}}_{\KS\Kp} - \mathcal{A}^{\textrm{det+prod}}_{\KS\Kp} - \mathcal{A}^{\textrm{\Kz}},
\end{eqnarray}
and similarly for \decay{\Bp}{\KS\pip}, except for the opposite sign of $\mathcal{A}^{\textrm{\Kz}}$, which is associated with the different flavor of \Kz in the two decays.

The detection and production asymmetries are determined using the \decay{\Bu}{\jpsi\Kp} control channel selected in the same fiducial volume as the signal. The combined detection and production asymmetry for charged kaons and \Bpm mesons is obtained by subtracting the known \CP asymmetry~\cite{PDG2024} from the measured raw asymmetry of the control channel. This is evaluated in bins of \Kp momentum and combined through a weighted average using the \decay{\Bp}{\KS\Kp} momentum spectrum, resulting in \mbox{$\mathcal{A}^{\rm det+prod}_{\KS\Kp}=(-1.02 \pm 0.32)\%$}. 

For the \decay{\Bu}{\KS\pip} decay, additional corrections are needed for the detection asymmetry difference between kaons and pions, and for \CP violation in the neutral kaon system. The combined asymmetry is
\begin{align}
    \mathcal{A}^{\rm det+prod}_{\KS\pip} &= \mathcal{A}^{\rm det+prod}_{\KS\Kp} - \mathcal{A}^{\Kp\pim} = \mathcal{A}^{\rm det+prod}_{\KS\Kp} - (\mathcal{A}^{\Kp\pim} + \mathcal{A}^{\Kz}) + \mathcal{A}^{\Kz},
\end{align}
where the combined asymmetry $(\mathcal{A}^{\Kp\pim}+\mathcal{A}^{\Kz})$ has previously been measured as a function of the charged kaon momentum~\cite{Davis:2310213}.
Potential effects from \CP violation in the neutral kaon system, either through $\Kz–\Kzb$ mixing~\cite{Grossman_2012} or \KL regeneration~\cite{Ko_2011}, are evaluated. The regeneration effect, where \KL interactions with detector material produce a \KS component, could contaminate the \decay{\Bu}{\KS h^+} sample with \decay{\Bp}{\KL h^+} decays. 
The resulting \Kz asymmetry is measured to be \mbox{$\mathcal{A}^{\Kz} = (-0.07 \pm 0.01)\%$} in both channels. The $K\pi$ asymmetry is determined to be \mbox{($\mathcal{A}^{\Kp\pim} + \mathcal{A}^{\Kz}) = (-0.64 \pm 0.01)\%$}. The final detection and production asymmetry for charged pion is \mbox{$\mathcal{A}^{\rm det+prod}_{\KS\pip}= (-0.31 \pm 0.32)\%$}.

% BR
The ratio of branching fractions is computed as
  \begin{align}
  \label{eq:MasterFormulaBR}
    \mathcal{R} \equiv \frac{\BR(\decay{\Bp}{\KS\Kp})}{\BR(\decay{\Bu}{\KS\pip})} &= \frac{N(\decay{\Bp}{\KS\Kp})}{N(\decay{\Bu}{\KS\pip})} \times \frac{\varepsilon(\decay{\Bu}{\KS\pip})}{\varepsilon(\decay{\Bp}{\KS\Kp})}, 
  \end{align}
where $\varepsilon$ denotes the total efficiency, including acceptance, reconstruction and selection effects. All efficiencies are derived from simulation after applying data-driven corrections. 

% Bc
The search for the \decay{\Bc}{\KS\Kp} decay employs a BDTG trained on simulated ${\decay{\Bc}{\KS\Kp}}$ decays as signal and mass-sideband candidates as background. A working point is chosen to maximize the figure of merit \mbox{$\varepsilon_{\text{BDTG}} / (2.5+\sqrt{N_{\rm B}})$}~\cite{Punzi:2003bu}, where $\varepsilon_{\text{BDTG}}$ is the signal efficiency of the BDTG requirement and $N_{\rm B}$ is the expected background yield in the \Bc signal mass region. The BDTG uses the same input variables as for the \decay{\Bp}{\KS\Kp} mode. No significant signal is observed. An upper limit on the branching fraction ratio multiplied by the \bquark-quark fragmentation probabilities, $f_{c}$ and $f_{u}$ corresponding to the hadronization fractions of a \bquark-quark into a \Bcp and a \Bp meson respectively, $\mathcal{R}^{'} \equiv \frac{f_{c}}{f_{u}}\cdot\frac{\BR(\Bcp \to \KS\Kp)}{\BR(\Bp \to \KS\pip)}$ is determined using the profile-likelihood method~\cite{Cowan:2010js,Schott:2012zb}. Systematic uncertainties are incorporated via a smeared likelihood-ratio function. The \Bcp signal yield is obtained from a single Gaussian fit to data with the mean fixed to the known \Bc mass~\cite{PDG2024} and the width taken from simulation. The combinatorial background is assumed to be a first-order polynomial function. The mass distribution and the fit projection are shown in the End~Matter.

%% uncertainties
The systematic uncertainties for the \ACP and $\mathcal{R}$ measurements are summarized in Table~\ref{tab:total error}. Systematic uncertainties from the fit model are evaluated by using alternative functions to describe the signal and background shapes, and by allowing \Bp and \Bm peak positions to vary independently within the baseline model.

\begin{table}
    \centering
    \caption{Systematic uncertainties on \ACP and $\mathcal{R}$ measurements. Absolute uncertainties are quoted for \ACP, while relative uncertainties are given for $\mathcal{R}$.}
    %\resizebox{\linewidth}{!}{
    \begin{tabular}{l|c|c|c}
    \hline
         &\multicolumn{2}{c|}{\CP asymmetry (\%)}& \multirow{2}{*}{$\mathcal{R}$ (\%)}\\
         \cline{2-3}
         & \decay{\Bp}{\KS\pip} &\decay{\Bp}{\KS\Kp}&\\
         \hline
         Fit model                          & 0.15  & 1.29  & 1.92 \\
         PID weighting                      & 0.01  & 0.01  & 0.35 \\
         $\mathcal{A}^{\textrm{det+prod}}$  & 0.32  & 0.32  & --  \\
         $\mathcal{A}^{\Kp\pim}+\mathcal{A}^{\Kz}$& --& 0.01  & --\\          
         $\mathcal{A}^{\textrm{\Kz}}$       & 0.01  & 0.01  &  --\\
         Trigger                         & 0.70  & 0.74  & 0.23  \\

         Simulation sample size               & --      & --       & 0.87 \\
         
         BDTG selection                     & 0.41  & 2.56 & --  \\   
         Mass veto selection                & 0.10  & 0.88 & 3.31  \\ 
         Tracking efficiency                   & --     &  --     & 0.01  \\
         \kaon/\pion interaction            & --     &  --     & 0.30  \\
         \hline 
         Total                                & 0.89 & 3.11 & 3.96 \\  
         \hline
    \end{tabular}
    %}
    \label{tab:total error}
\end{table}

Systematic uncertainties associated with the PID corrections applied to the simulation account for effects from the PID efficiency binning scheme, calibration sample statistics, PID response modelling. These uncertainties are combined in quadrature. 
The total systematic uncertainty from detector and production asymmetries is determined to be 0.32\%, including a 0.30\% contribution from the uncertainty on $\ACP(\decay{\Bu}{\jpsi\Kp})$~\cite{PDG2024}. 
Trigger asymmetries are incorporated in the overall production and detection corrections, which are estimated from \decay{\Bu}{\jpsi\Kp} candidates, with only the associated uncertainties retained in this analysis using a data-driven approach.

Systematic uncertainties related to selection efficiencies are evaluated, including the limited size of the simulated samples. In addition, the stability of the measured observables is tested by varying the selection criteria.
The systematic uncertainty associated with the BDTG selection is evaluated by repeating the full measurement while varying the BDTG requirement around the nominal working point. The resulting variation in the fitted is assigned as the systematic uncertainty. No uncertainty is assigned to the branching-fraction ratio, while uncertainties of 0.41\% and 2.56\% are obtained for the \KS\pip and \KS\Kp modes, respectively; the larger value for is mainly driven by the smaller effective signal yield, making the result more sensitive to statistical fluctuations than to changes in the BDTG efficiency. 
For the \decay{\Bu}{\KS\pip} and \decay{\Bp}{\KS\Kp} samples, an additional uncertainty is assigned by varying the mass veto on the \proton\pim invariant mass from $\pm 6 \mevcc$ to $\pm 10 \mevcc$ of the known \Lz mass. Furthermore, for the \decay{\Bp}{\KS\Kp} samples only, two mass vetoes are varied to evaluate the associated systematic uncertainties. The first is applied to the \Kp\pim invariant mass, where the veto window around the known \Dz mass is changed from $\pm15\mevcc$ to $\pm20\mevcc$. The second is applied to the \Bp candidate mass, used to suppress background from misidentified \decay{\Bp}{\KS\pip} decays, where the baseline requirement of $\pm45\mevcc$ is varied to $\pm40\mevcc$ and $\pm50\mevcc$. The systematic uncertainties from these variations on \ACP are 0.10\% and 0.88\% for the \KS\pip and \KS\Kp modes, respectively, and 3.31\% on the branching fraction ratio.

Tracking efficiency corrections are applied to simulated candidates as a function of \textit{p} and $\eta$ for each final-state particle, with the statistical uncertainty from the data calibration sample included. 
Additional systematic uncertainties arise from imperfect modeling of kaon and pion interactions with detector material in the simulation. The detection efficiency uncertainties from material interactions are estimated to be 1.1\% for kaons and 1.4\% for pions, determined by varying the detector material budget in the simulation by approximately $\pm$10\%~\cite{DeCian:1402577}. These uncertainties are treated as fully correlated, resulting in a relative uncertainty of 0.3\% for the ratio of pion to kaon efficiencies.

%% results
The \CP asymmetries are determined to be
\begin{eqnarray}
\ACP({\decay{\Bu}{\KS\pip}}) &=& -0.028\pm 0.009 \pm 0.009,
\nonumber \\
\ACP({\decay{\Bp}{\KS\Kp}}) &=& \phantom{-}0.118\pm 0.062 \pm 0.031 \nonumber,
\end{eqnarray}
where the first uncertainty is statistical and the second systematic, representing about a factor-of-two improvement in precision compared to the Run~1 results~\cite{LHCb-PAPER-2013-034}. The combined statistical and systematic correlation between the asymmetries is estimated to be $-0.10$.
The measured \ACP(\decay{\Bu}{\KS\pip}) value shows good consistency with previous results from \babar~\cite{BaBar:2006enb}, \belle~\cite{Belle:2012dmz}, and \lhcb in Run~1~\cite{LHCb-PAPER-2013-034}, while exhibiting a $2.3\sigma$ difference with the most recent \belletwo measurement~\cite{Belle-II:2023ksq}. The new \lhcb result differs from theoretical predictions by $2.4\sigma$ from QCDF calculations~\cite{Cheng:2009cn,Cheng:2014rfa} and by $2.1\sigma$ from the pQCD prediction~\cite{Chai:2022ptk}.
Similarly, the measured \ACP(\decay{\Bp}{\KS\Kp}) value shows differences of $2.5\sigma$ and $1.4\sigma$ with respect to QCDF and pQCD predictions, respectively. These indicate potential challenges in the theoretical description of hadronic effects in these decays.

The measured branching fraction ratio is 
\begin{eqnarray}
\nonumber
\frac{\BR(\decay{\Bp}{\KS\Kp})}{\BR(\decay{\Bu}{\KS\pip})} = 0.055\pm 0.004 \pm 0.002,
\end{eqnarray}
showing good agreement with both the previous \lhcb measurement~\cite{LHCb-PAPER-2013-034} and theoretical predictions~\cite{BurgosMarcos:2025xja}. 

The upper limit on the branching fraction ratio multiplied by the \bquark-quark fragmentation probabilities between the \decay{\Bc}{\KS\Kp} and \decay{\Bu}{\KS\pip} decay modes is determined at the 90\% (95\%) confidence level (CL) as 
\begin{eqnarray}
\nonumber
\frac{f_c}{f_u}\cdot\frac{\BR({\decay{\Bc}{\KS\Kp}})}{\BR({\decay{\Bu}{\KS\pip}})} < 0.015 \, (0.016) \textrm{ at 90\% (95\%) CL}.
\end{eqnarray}

In summary, all measurements presented in this Letter are consistent with SM predictions, yet they provide the most precise experimental constraints to date on \CP violation in these charmless hadronic \B decays. The precision achieved for $\ACP({\decay{\Bu}{\KS\pip}})$ with a total uncertainty of 0.013, a reduction by a factor of two compared to the previous best measurement, establishes a new benchmark for a theoretically clean null-test channel in the $b \to s$ sector. 
Similarly, the measurement of \ACP({\decay{\Bu}{\KS\Kp}}) reaches a precision that begins to discriminate between leading theoretical approaches, providing unique insight into hadronic dynamics.
No significant signal is observed for the decay {\decay{\Bc}{\KS\Kp}}, and the resulting upper limit can be used to constrain the poorly understood weak annihilation contributions in many hadronic $b$-meson decays.
%The upper limit set on {\decay{\Bc}{\KS\Kp}} can be used to constrain weak annihilation processes. 
Collectively, these results not only set new standards for precision but also establish a foundational dataset for future global analyses aimed at resolving persistent puzzles in flavor physics.
With the increased statistic from the upgraded \lhcb ~\cite{LHCb-DP-2022-002} and \belletwo experiments, future investigations will build upon this precision frontier, further enhancing the sensitivity to potential
deviations from the SM.
\clearpage
%\input{LHCb-latex-template/latest/latex/mywordcount}
% Do not include this in any draft (just for information in the template)
%\input{acknowledgements_intro}
% Comment this in for paper drafts; do not include this in analysis note, conference and figure reports
\section*{Acknowledgements}
%
% These Acknowledgements valid from 12/09/25
%
\noindent We express our gratitude to our colleagues in the CERN
accelerator departments for the excellent performance of the LHC. We
thank the technical and administrative staff at the LHCb
institutes.
We acknowledge support from CERN and from the national agencies:
ARC (Australia);
CAPES, CNPq, FAPERJ and FINEP (Brazil); 
MOST and NSFC (China); 
CNRS/IN2P3 (France); 
BMFTR, DFG and MPG (Germany);
INFN (Italy); 
NWO (Netherlands); 
MNiSW and NCN (Poland); 
MCID/IFA (Romania); 
%MSHE (Russia); 
MICIU and AEI (Spain);
SNSF and SER (Switzerland); 
NASU (Ukraine); 
STFC (United Kingdom); 
DOE NP and NSF (USA).
%%%%%%%%%%%%%%%%%%%%%%%%%%%%%%%%%%%%%%%%%%%%%
We acknowledge the computing resources that are provided by ARDC (Australia), 
CBPF (Brazil),
CERN, 
IHEP and LZU (China),
IN2P3 (France), 
KIT and DESY (Germany), 
INFN (Italy), 
SURF (Netherlands),
Polish WLCG (Poland),
IFIN-HH (Romania), 
%RRCKI and Yandex LLC (Russia), 
PIC (Spain), CSCS (Switzerland), 
and GridPP (United Kingdom).
%%%%%%%%%%%%%%%%%%%%%%%%%%%%%%%%%%%%%%%%%%
We are indebted to the communities behind the multiple open-source
software packages on which we depend.
%%%%%%%%%%%%%%%%%%%%%%%%%%%%%%%%%%%%%%%%%%
Individual groups or members have received support from
%ARC and ARDC (Australia); % moved to national 16/01
Key Research Program of Frontier Sciences of CAS, CAS PIFI, CAS CCEPP, 
%Fundamental Research Funds for the Central Universities,  and Sci.\
%\& Tech.\ Program of Guangzhou (China); Removed 24/11/25
Minciencias (Colombia);
EPLANET, Marie Sk\l{}odowska-Curie Actions, ERC and NextGenerationEU (European Union);
A*MIDEX, ANR, IPhU and Labex P2IO, and R\'{e}gion Auvergne-Rh\^{o}ne-Alpes (France);
%RFBR, RSF and Yandex LLC (Russia);
Alexander-von-Humboldt Foundation (Germany);
ICSC (Italy); 
%GVA, XuntaGal, GENCAT, Inditex, InTalent and Prog.~Atracci\'on Talento, CM (Spain);
Severo Ochoa and Mar\'ia de Maeztu Units of Excellence, GVA, XuntaGal, GENCAT, InTalent-Inditex and Prog.~Atracci\'on Talento CM (Spain);
SRC (Sweden);
the Leverhulme Trust, the Royal Society and UKRI (United Kingdom).

%\input{supplementary}

%\input{appendix}

% This should be taken out in the final paper
\clearpage

\section*{End Matter}
\label{sec:Supplementary-App}

\section{Study of \boldmath{\BcToKSK} mass distribution}
\begin{figure}[!b]
    \centering
    \includegraphics[width=0.7\linewidth]{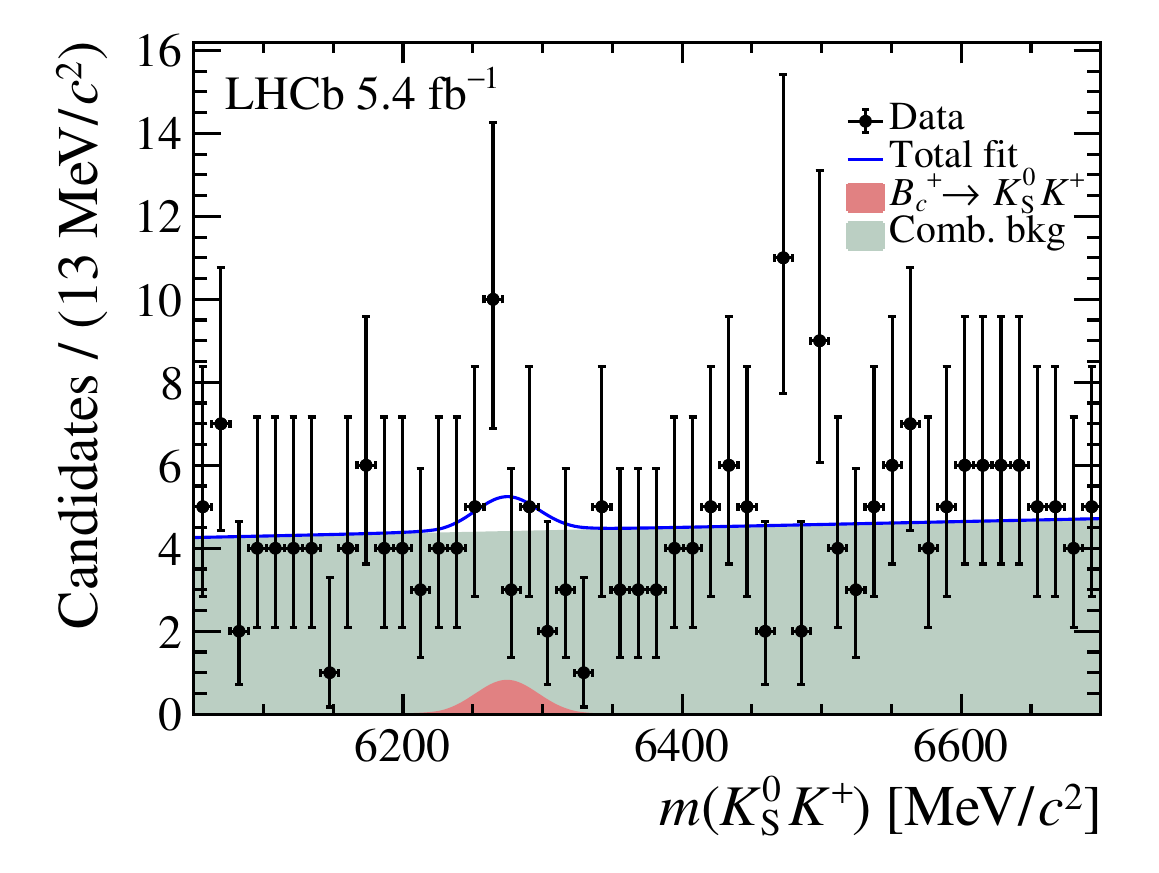}
    \caption{Mass distribution of selected $B_c^{+} \to K^0_{\mathrm{S}} K^{+}$ candidates. The data points with error bars are shown together with the result of the fit (solid curve). 
    }
    \label{fig2}
\end{figure}

The mass distribution of \BcToKSK candidates is shown in Fig.~\ref{fig2}. The signal significance is determined to be $1.0$ standard deviation. Systematic studies were performed by varying the BDTG selection requirement and fixing the mean value and mass resolution of \Bc signal to the known value and simulation predictions, respectively. The significance determination is found to be stable under these variations. 

\section{Comparison with previous results and theory}
\begin{figure}[!tb]
    \centering
    \includegraphics[width=0.62\linewidth]{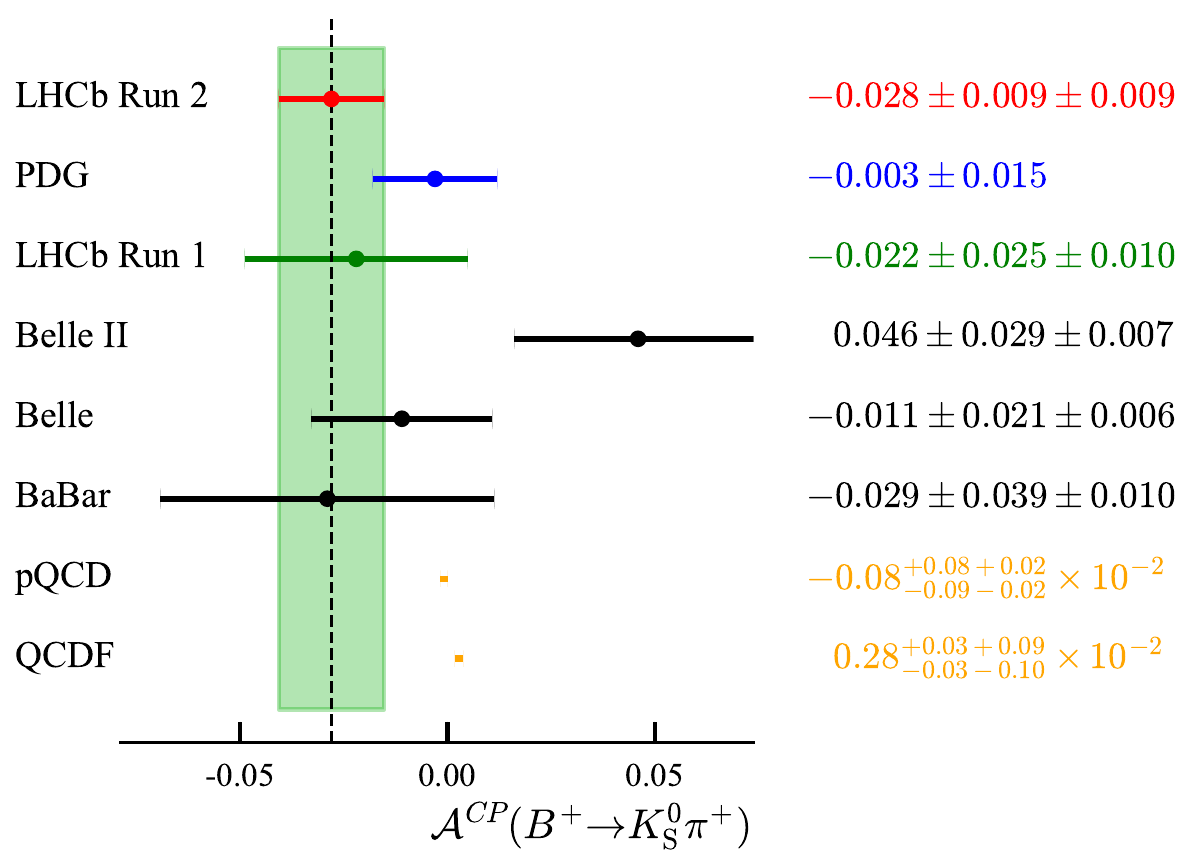}
    \vspace{1em}
    \includegraphics[width=0.62\linewidth]{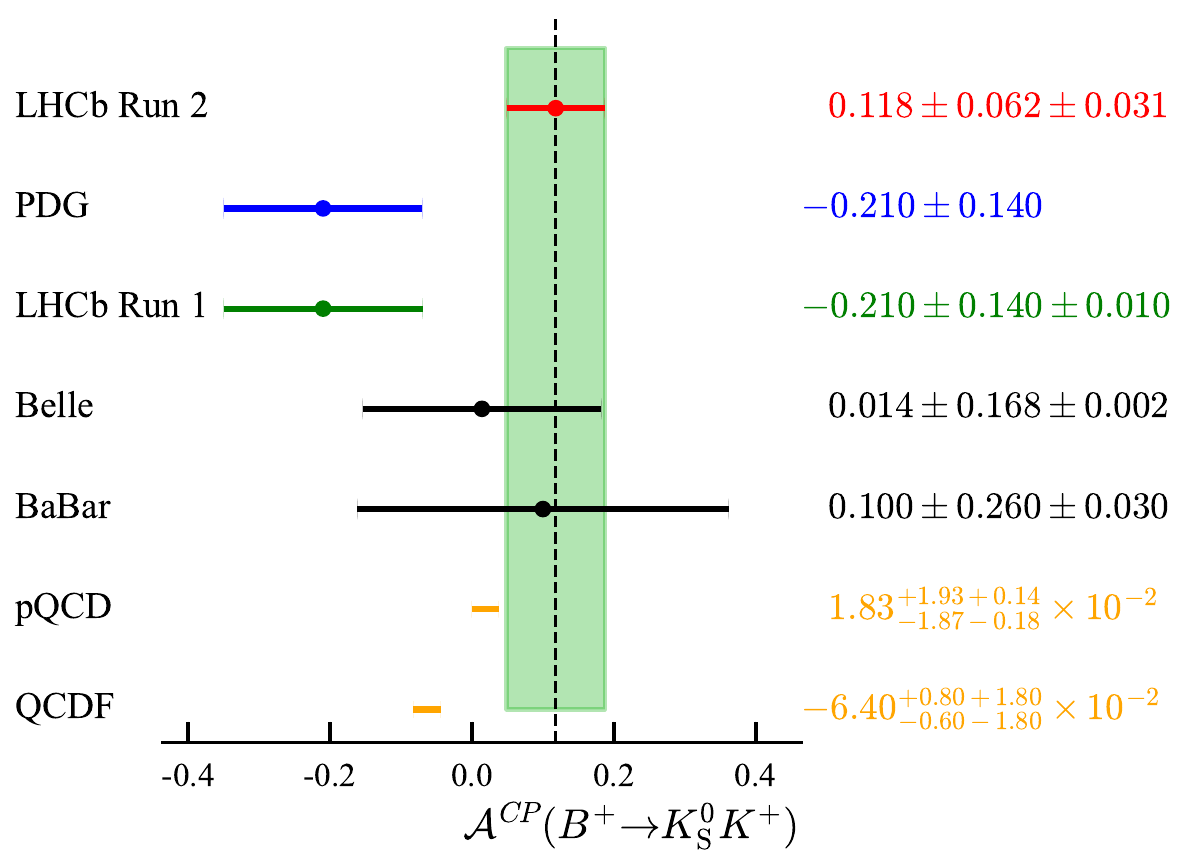}
    \vspace{1em}
    \includegraphics[width=0.62\linewidth]{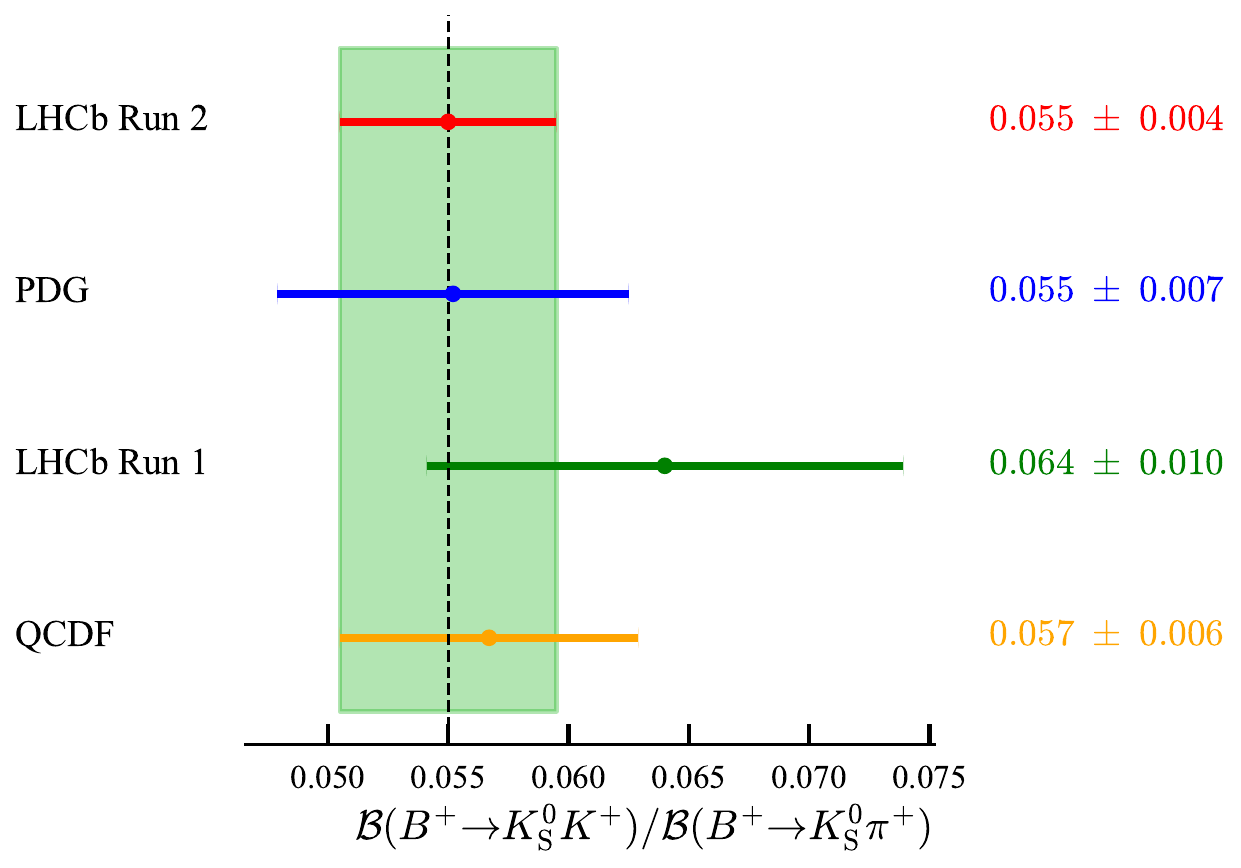}
    \caption{Comparison of branching fraction ratios measured in this paper with results from 
    different experiments~\cite{BaBar:2006enb, Belle:2012dmz, LHCb-PAPER-2013-034, Belle-II:2023ksq,PDG2024} and theoretical predictions~\cite{Cheng:2009cn,Cheng:2014rfa,Chai:2022ptk,BurgosMarcos:2025xja}.}
    \label{fig:compare_results}
\end{figure}

Figure~\ref{fig:compare_results} shows comparisons of the measured \CP asymmetries and branching fraction ratio with previous experimental results and theoretical predictions.

\clearpage

\addcontentsline{toc}{section}{References}
%\setboolean{inbibliography}{true}
\bibliographystyle{LHCb}
\bibliography{main,standard,LHCb-PAPER,LHCb-CONF,LHCb-DP,LHCb-TDR,mybib}

\newpage
% LHCb collaboration author list
% Data extracted on April 23rd, 2026 at 10:53am for paper reference LHCb-PAPER-2025-049
\centerline
{\large\bf LHCb collaboration}
\begin
{flushleft}
\small
R.~Aaij$^{38}$\lhcborcid{0000-0003-0533-1952},
A.S.W.~Abdelmotteleb$^{58}$\lhcborcid{0000-0001-7905-0542},
C.~Abellan~Beteta$^{52}$\lhcborcid{0009-0009-0869-6798},
F.~Abudin\'en$^{58}$\lhcborcid{0000-0002-6737-3528},
T.~Ackernley$^{62}$\lhcborcid{0000-0002-5951-3498},
A.A.~Adefisoye$^{70}$\lhcborcid{0000-0003-2448-1550},
B.~Adeva$^{48}$\lhcborcid{0000-0001-9756-3712},
M.~Adinolfi$^{56}$\lhcborcid{0000-0002-1326-1264},
P.~Adlarson$^{86}$\lhcborcid{0000-0001-6280-3851},
C.~Agapopoulou$^{14}$\lhcborcid{0000-0002-2368-0147}
% ...

{\footnotesize \it

% ...
$^{14}$Universit{\'e} Paris-Saclay, CNRS/IN2P3, IJCLab, Orsay, France\\
% ...
$^{38}$Nikhef National Institute for Subatomic Physics, Amsterdam, Netherlands\\
% ...
$^{48}$Instituto Galego de F{\'\i}sica de Altas Enerx{\'\i}as (IGFAE), Universidade de Santiago de Compostela, Santiago de Compostela, Spain\\
% ...
$^{52}$Physik-Institut, Universit{\"a}t Z{\"u}rich, Z{\"u}rich, Switzerland\\
% ...
$^{56}$H.H. Wills Physics Laboratory, University of Bristol, Bristol, United Kingdom\\
% ...
$^{58}$Department of Physics, University of Warwick, Coventry, United Kingdom\\
% ...
$^{61}$School of Physics and Astronomy, University of Glasgow, Glasgow, United Kingdom\\
$^{62}$Oliver Lodge Laboratory, University of Liverpool, Liverpool, United Kingdom\\
% ...
$^{70}$Syracuse University, Syracuse, NY, United States\\
% ...
$^{86}$Department of Physics and Astronomy, Uppsala University, Uppsala, Sweden, associated to $^{61}$\\
% ...
}
\end{flushleft}
%\input{~/Downloads/Authorship_LHCb-SD-2024-001-grouped-internal-use-only-do-not-publish-2.tex}

%PRL Justification:

%The study of \CP violation in charmless \B decays provides a powerful probe of the Standard Model (SM) and sensitive tests of potential new physics. While \CP asymmetries in \BuToKSPi and \BuToKSK decays are predicted to be small in the SM, they are highly sensitive to contributions from beyond-SM processes. Our precision measurements of $\ACP(\BuToKSPi)=\ACPKPiResult$ and $\ACP(\BuToKSK)=\ACPKKResult$, representing the most stringent tests to date, constrain hadronic uncertainties and challenge theoretical predictions. The simultaneous search for \BcToKSK, rare in the SM but enhanced in certain new physics scenarios, yields the world’s most sensitive upper limit. These results lay the foundation for future \CP-violation studies at high-luminosity colliders.

\end{document}